\documentclass[aps,prapplied,reprint,longbibliography,superscriptaddress]{revtex4-1}

\usepackage{color,amsmath,amssymb,graphicx}
\usepackage{threeparttable}
\usepackage[caption=false]{subfig}
\usepackage{graphicx}

\begin{document}


\title{Experimental demonstration and in-depth investigation of analytically designed anomalous reflection metagratings}


\author{Oshri Rabinovich}
\affiliation{Andrew and Erna Viterbi Faculty of Electrical Engineering, Technion - Israel Institute of Technology, Haifa 3200003, Israel}
\affiliation{Rafael, Advanced Defense Systems, Ltd.\\
Haifa 31021, Israel}

\author{Ilan Kaplon}
\affiliation{Rafael, Advanced Defense Systems, Ltd.\\
Haifa 31021, Israel}

\author{Jochanan Reis}
\affiliation{Rafael, Advanced Defense Systems, Ltd.\\
Haifa 31021, Israel}

\author{Ariel Epstein}
\email[]{epsteina@ee.technion.ac.il}
\affiliation{Andrew and Erna Viterbi Faculty of Electrical Engineering, Technion - Israel Institute of Technology, Haifa 3200003, Israel}


\date{\today}

\begin{abstract}

We present the design, fabrication and experimental investigation of a printed circuit board (PCB) metagrating (MG) for perfect anomalous reflection. 
The design follows our previously developed analytical formalism, 
resulting in a single-element MG capable of unitary coupling of the incident wave to the specified (first order) Floquet-Bloch (FB) mode while suppressing the specular reflection. We characterize the MG performance experimentally using a bistatic scattering pattern measurement, relying on an original beam-power integration approach for accurate evaluation of the coupling to the various modes across a wide frequency range. The results show that highly-efficient wide-angle anomalous reflection is achieved, as predicted by the theory, with a relatively broadband response. In addition, the MG is found to perform well when illuminated from different angles, acting as a multichannel reflector, and to scatter efficiently also to higher-order FB modes at other frequencies, exhibiting multifunctional capabilities.
Importantly, the merits of the introduced beam-integration approach, namely, its improved resilience to measurement inaccuracies or noise effects, and the implicit accommodation of different effective aperture sizes, are emphasized, highly relevant in view of the numerous recent experimental reports on anomalous reflection metasurfaces. 
Finally, we discuss the source for losses associated with the MG; interestingly, we show that these correlate well with edge diffraction effects, rather than (commonly assumed) power dissipation. These experimental results verify our theoretical synthesis scheme, showing that highly-efficient anomalous reflection is achievable with a realistic fabricated MG, demonstrating the practical applicability and potential mutifunctionality of analytically designed MGs for future wavefront manipulating devices.

\end{abstract}

\pacs{}

\maketitle

\section{Introduction}
\label{sec:introduction}
Metagratings (MGs), periodic structures comprised of one or a few polarizable meta-atoms per period, are attracting considerable attention lately 
\cite{khorasaninejad2015broadband,sell2017large,fan2018perfect,ra2018reconfigurable,ra2017meta,wong2018perfect,wong2018binary}
, due to their ability to manipulate beams with very high efficiencies, usually difficult to achieve with gradient metasurfaces (MS) \cite{yu2011light,estakhri2016recent}. Furthremore, they can also overcome realization difficulties related to conventional Huygens' and bianisotropic MSs, as the latter require microscopic design of multiple closely-packed meta-atoms per period, typically corresponding to a significantly larger number of design degrees of freedom \cite{kuester2003averaged,pfeiffer2013metamaterial,pfeiffer2014bianisotropic,epstein2016huygens,epstein2016arbitrary,epstein2016synthesis,asadchy2016perfect,lavigne2018susceptibility,chen2018theory}.

In particular, the concept of metagratings for anomalous reflection was rigorously introduced and formulated in a recent paper by Ra'di et al. \cite{ra2017meta} using a repeating magnetically-polarizable particle (conducting loop) above a perfect electric conductor (PEC) ground plane. 
In the wide-angle reflection scenarios considered therein, the transverse momentum difference between the incident wave and the desirable reflected wave defines a grating periodicity $\Lambda$ that allows only two Floquet-Bloch (FB) modes to propagate: the specular (fundamental) and anomalous (first order) reflection modes; the rest of the modes are evanescent \cite{russell1986optics}.
Subsequently, the authors showed that enforcing two physical constraints, namely, cancellation of specular reflection and global power conservation, allows synthesis of perfect anomalous reflection MGs using only two degrees of freedom. First, based on these two constraints, a nonlinear condition was analytically formulated, resolving the required MG-PEC separation distance $h$; next, once this parameter was set, the meta-atom geometry was tuned to optimize the coupling to the desirable first-order FB mode. This yielded an appealing theoretical scheme for designing simple yet highly efficient beam deflectors; nonetheless, it has yet to be experimentally tested. 

Lately, analytical models for electrically-polarizable MGs were developed, relying on arrays of dielectric rods \cite{chalabi2017efficient} or capacitively-loaded wires \cite{epstein2017unveiling} for achieving wide-angle anomalous reflection and beam splitting. In fact, the theoretical framework was augmented in \cite{epstein2017unveiling} to allow \emph{complete} semianalytical design of the MG, up to the conductor geometry of the printed capacitors. Notwithstanding, although these reports marked a significant advance towards an experimental demonstration by utilizing meta-atom geometries more compatible with common 2D fabrication techniques, they still considered the MGs to be 
\emph{suspended in air} above the ground plane, thus not suitable for practical realization.


\begin{figure*}[t]
\includegraphics[width=13cm]{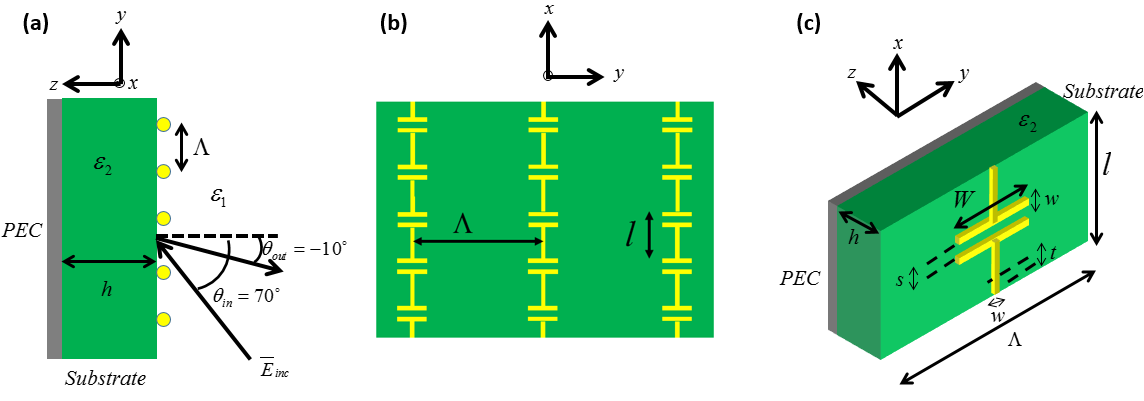}
\caption{Physical configuration of the PCB MG realized and characterized herein. (a) Side view of the MG configuration  
(b) Top view of the MG, featuring the capacitively-loaded wires. The printed capacitor unit cells are repeating with a subwavelength period $l\ll\lambda$ along the $x$ axis.
comprised of loaded strips with period $\Lambda$, defined on a dielectric substrate backed by a metal (PEC) layer. 
(c) The MG printed capacitor unit cell, featuring a single degree of freedom, namely, the capacitor width $W$. The trace separation is $s$, its width is $w$, and the copper thickness is $t.$}
\label{Unit_Cell}
\end{figure*}

Aiming at such a realization, we have extended in a recent paper \cite{rabinovich2018analytical} the model of \cite{epstein2017unveiling} as to consider a structure fully-compatible with printed circuit board (PCB) implementation, namely, having a dielectric substrate between the MG and the PEC (Fig. \ref{Unit_Cell}). This extension required accounting for multiple reflections within the Fabry-P\'{e}rot etalon formed by the PEC-backed dielectric slab, introducing modal reflection coefficients into the formulation; and incorporation of an effective dielectric constant when deriving the printed capacitor geometry, positioned at the interface between air and the dielectric \cite{rabinovich2018analytical}.
The realistic analytical model produces detailed fabrication-ready design specifications of PCB MGs for perfect anomalous reflection, obtained without requiring even a single simulation in full-wave solvers.  
These results naturally pave the path to experimental demonstration of single-element MGs designed following this analytical scheme, yet to be carried out.

With the recent advances in the research of MSs for anomalous reflection \cite{epstein2016synthesis,asadchy2016perfect}, experiments conducted in order to characterize these structures and verify their functionality at microwave frequencies were reported \cite{diaz2017generalized,wong2018perfect,asadchy2017flat,wong2018binary}. In order to measure both the specular and anomalous reflections, the experimental setup often consisted of a single rotating element: either the MS was rotated while the transmitter and receiver antennas were fixed \cite{diaz2017generalized}, or the receiver antenna was rotated while the transmitter and MS were fixed \cite{wong2018perfect}. In both cases, the efficiency of anomalous reflection was deduced via comparison to the scattering from a metallic plate of the same size. 
Furthermore, the power coupled to the various FB modes was inferred from the peaks of the corresponding beams (a single measurement point for each propagating FB mode); hence, to achieve meaningful results, the authors had to factor out the differences in effective aperture sizes at different observation angles \cite{sievenpiper2005forward}.

In this paper, we present the design, fabrication and experimental characterization of a MG that is capable of reflecting an incoming wave into a prescribed angle (anomalous reflection) with very high efficiency, validating our analytical model \cite{rabinovich2018analytical}. In contrast to previous realizations of this functionality, our MG is exclusively designed, up to the complete PCB layout, following a rigorous analytical scheme \cite{rabinovich2018analytical}, featuring only a single element with a single degree of freedom per period. 
The in-depth investigation we conduct examines the angular response of the MG at various frequencies, revealing its potential as a multifunctional device, and probes the MG performance when illuminated from non-designated angles, indicating its broad acceptance angle. Overall, our results verify experimentally that MGs can generally implement highly-efficient anomalous reflection over a relatively wide range of frequencies and angles of incidence, as pointed out in \cite{ra2017meta,epstein2017unveiling,rabinovich2018analytical,wong2018perfect}.

In addition to providing experimental proof to the practical viability of analytically designed MGs, we suggest here an original approach to the characterization of anomalous reflection from planar scattering devices such as MSs and MGs, differing from the common methodologies \cite{diaz2017generalized,wong2018perfect,asadchy2017flat,wong2018binary}. Our proposed experimental scheme relies on measurements of the bistatic scattering from the MG, utilizing a fine angular resolution over the half space $z>0$. Similar to \cite{wong2018perfect}, we illuminate both the MG and the reference metal plate from the designated angle of incidence and record the scattering pattern in all directions. However, in contrast to these previous experimental reports, we employ a beam-integration approach to characterize the coupling efficiency to the various FB modes. Specifically, we identify the scattered beam corresponding to each one of the propagating FB modes, and integrate the pattern over the angular range associated with it to estimate the power coupled to that mode. 

The proposed approach has two appealing features. First, it relies on an integrated set of measurement points, thus reducing the impact of measurement noise. Second, integrating over the entire beam ensures that all of the radiated power associated with it is properly accounted for, avoiding the need to explicitly take into consideration the differences in effective aperture sizes when comparing beams at different angles \cite{diaz2017generalized,wong2018perfect,asadchy2017flat,wong2018binary} (such a calibration may be especially challenging when the measurements are not entirely in the far field).
In view of the growing interest in MSs and MGs implementing controlled reflection, the presented alternative approach, which may lead to improved experimental accuracy and can be applied to other scattering scenarios as well, forms a valuable contribution of this work.

Finally, we quantify the losses in the fabricated MG, and investigate their origin. Surprisingly, our combined experimental and theoretical evidence implies that the observed losses are \emph{not} related to dissipation in the dielectric substrate and metallic traces, but rather stems from the finite size of the characterized sample, leading to increased edge diffraction, which is interpreted as a loss mechanism in our setup. These results are important for two reasons: first, they establish the MG as a low-absorption device, as predicted by realistic simulations \cite{rabinovich2018analytical,epstein2017unveiling}; second, they highlight a loss mechanism that is hardly considered in MG and MS experimental work, although all measurement setups involve finite devices that could be prone to the same phenomenon.

Overall, the comprehensive experimental work reported herein establishes analytically-designed MGs as a viable alternative to MSs for the realization of planar, highly-efficient, wide-angle beam manipulating devices, performing very well in the presence of realistic fabrication inaccuracies, conductor and dielectric losses. Together with the demonstrated potential for multifunctionality, these results provide further support and motivation for future exploration of this emerging concept. Equally important, the presented beam-integration approach for characterizing coupling efficiencies and the new light shed on the origin of apparent losses in these devices are expected to enable proper interpretation of the measured results not only for MGs, but for general scattering configurations as well.

\section{Theory, design, and experiment}
\label{sec:theory}

We follow the analytical scheme of \cite{rabinovich2018analytical} for the design of a MG prototype at K band that fully couples an incoming plane wave at an angle of $\theta_\mathrm{in}=70^{\circ}$ towards the non-specular direction $\theta_\mathrm{out}=-10^{\circ}$ [Fig. \ref{Unit_Cell}(a)]. 
The MG is comprised of capacitively-loaded wires defined on a metal-backed dielectric substrate of thickness $h$, corresponding to a standard single-layer PCB layout [Fig. \ref{Unit_Cell}(b) and (c)]. The periodicity $\Lambda$ along the $y$ axis facilitates the coupling between the incident and reflected wave momenta, thus is set as $\Lambda=\lambda/\left|\sin\theta_\mathrm{out}-\sin\theta_\mathrm{in}\right|$, typically comparable with the free-space wavelength $\lambda$ of the excitation fields. On the other hand, the periodicity $l$ along the $x$ axis is chosen to be much smaller, $l=\lambda/10\ll\lambda$, such that we can treat the closely-spaced capacitors as a uniform distributed impedance loading of the current-carrying wires \cite{ikonen2007modeling,ra2017meta,epstein2017unveiling}.

The full details regarding the synthesis procedure can be found in \cite{rabinovich2018analytical}; for completeness, we briefly repeat here the main formulation results, as applicable to the prototype designed herein. For given substrate permittivity $\varepsilon_2$ and frequency of operation $f=c/\lambda$ ($c$ is the speed of light in vacuum), the procedure focuses on setting the two degrees of freedom of the MG structure, namely, the substrate thickness $h$ and the capacitor width $W$ [Fig. \ref{Unit_Cell}(c)], such that full coupling of the incident wave towards the prescribed $\theta_\mathrm{out}$ would be achieved. In the first step, we assess the optimal $h$ by imposing the two design constraints, namely, elimination of specular reflection and power conservation, on the modal field expressions, yielding the condition for perfect anomalous reflection, reading \cite{rabinovich2018analytical}
\begin{equation}
\label{eq:final_condition}
\rho=\frac{\cos \theta_\mathrm{out}}{\cos \theta_\mathrm{in}}-\frac{\left|1+R_{-1}(h)\right|^{2}}{\left|1+R_{0}(h)\right|^{2}}=0,
\end{equation}
where $R_{m}=\frac{j\gamma_{m}\tan\beta_{m,2}h-1}{j\gamma_{m}\tan\beta_{m,2}h+1}$ is the reflection coefficient of the $m$th FB mode, $\gamma_{m}=\frac{\eta_{2}k_{2}\beta_{m,1}}{\eta_{1}k_{1}\beta_{m,2}}$ is the $m$th mode wave-impedance ratio, and $\beta_{m,p}=\sqrt{k_{p}^{2}-(k_{1}\sin\theta_\mathrm{in}+\frac{2\pi m}{\Lambda})^{2}}$ is the longitudinal wavenumber. The index $p$ stands for the medium in which the parameters are evaluated, with $p=1$ being air and $p=2$ being the dielectric substrate; correspondingly, $\eta_{p}$ and $k_{p}$ are the wave impedance and wave number in medium $p$. 

We note that the periodicity $\Lambda$ dictates the propagation angle of the $m$th FB mode scattered off the MG via the parameter $\beta_{m,1}$. Following the guidelines of \cite{rabinovich2018analytical,ra2017meta,epstein2017unveiling}, it can be verified that the period defined at the beginning of this section forms a FB series in which the $m=-1$ mode coincides with the desirable anomalous reflection toward $\theta_\mathrm{out}$, and all modes other than the ones corresponding to specular and anomalous reflections are evanescent ($\Im\{\beta_{m,1}\}<0, \forall m\neq0,-1$). For this reason, demanding that the specular reflection would vanish via Eq. \eqref{eq:final_condition} guarantees exclusive coupling of the real power to the $m=-1$ mode, as no other radiation channels are available.

Once the substrate thickness $h$ minimizing $|\rho|$ of Eq. \eqref{eq:final_condition} for the given $\theta_\mathrm{in}$ and $\theta_\mathrm{out}$ is found, we proceed to setting the MG's second degree of freedom, the capacitor width $W$, assessed from the load capacitance $C$ required to establish the suitable grid currents to sustain the scattering phenomena. Indeed, this capacitance is resolved by invoking Ohm's law with these required currents and the total fields acting on the wire shell 
\cite{tretyakov2003analytical,rabinovich2018analytical}. This leads to the following expression for $C$
\begin{align}
\label{eq:final_load_impedance}
\dfrac{1}{2\pi lfC}&\!=\!-\frac{\eta_{1}\left|1+R_{-1}\right|^{2}}{4\Lambda \cos \theta_\mathrm{out}}\!\left[\frac{1}{\gamma_{0}\tan \beta_{0,2}h}+\frac{1}{\gamma_{-1}\tan \beta_{-1,2}h}\right] \nonumber \\
&+\frac{k_{1}\eta_{1}}{2\pi}\left(\frac{1}{2}+\log \frac{\pi w}{2\Lambda}\right) \nonumber \\
&-\frac{\eta_{1}}{\Lambda}\sum_{\substack{m=-\infty \\ m\neq 0,-1}}^{\infty}\left[\frac{k_{1}(1+R_{m})}{2\alpha_{m,1}}-\frac{k_{1}\Lambda}{4\pi}\frac{1}{\left| m \right|}\right],
\end{align}
where $w$ is the width of the wire trace [Fig. \ref{Unit_Cell}(c)], and we define $\alpha_{m,p}=-j\beta_{m,p}$ ; substituting the optimal $h$ found earlier allows evaluation of the required load capacitance $C$ to implement the desirable functionality. Finally, when the capacitance is realized as a printed capacitor with identical trace width and separation $w=s$ [Fig. \ref{Unit_Cell}(b)], the capacitor width $W$ can be retrieved from $C$ using $W\approx 2.85K_{\mathrm{corr}}C/\varepsilon_{\mathrm{eff}}[\mathrm{mm}/\mathrm{fF}]$, where $K_\mathrm{corr}$ is a frequency dependent correction factor, and the average dielectric constant is $\varepsilon_{\mathrm{eff}}=\left(\varepsilon_1+\varepsilon_2\right)/\left(2\varepsilon_0\right)$ \cite{guptamicrostrip,rabinovich2018analytical,epstein2017unveiling}.

\begin{figure}[t]
\centering
\includegraphics[width=3.0in]{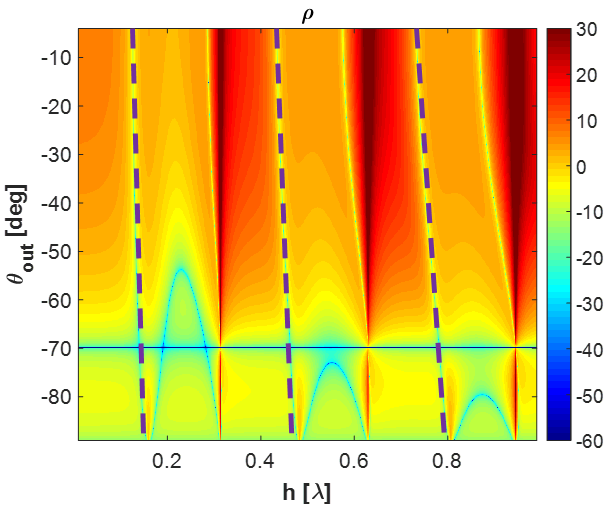}
\caption{Deviation from the perfect anomalous reflection condition $\left|\rho\right|$ of Eq. \eqref{eq:final_condition} (in dB scale), as a function of the substrate thickness $h$ and the reflection angle $\theta_\mathrm{out}$, for the considered MG prototype configuration ($\theta_\mathrm{in}=70^{\circ}$). The dashed purple vertical lines indicate the solution branches where the deviation $\left|\rho\right|$ is minimal. We choose the first branch as a preferable solution (see discussion in \cite{rabinovich2018analytical}).}
\label{Fig:Theta_out_h}
\end{figure}

\begin{figure}[t]
\centering
\includegraphics[width=3.0in]{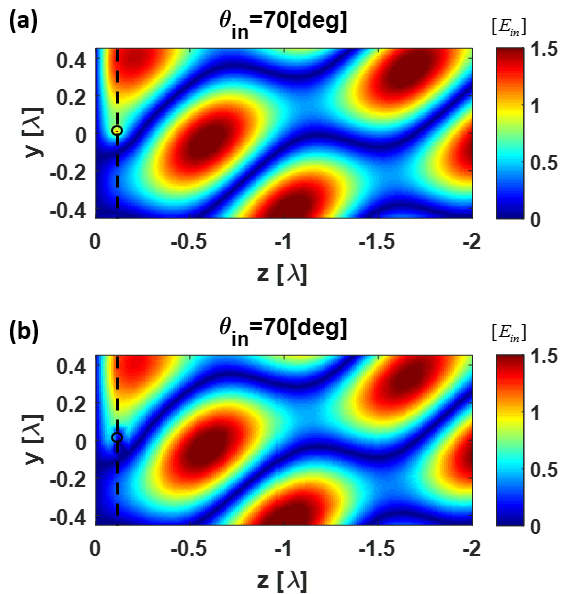}
\caption{Electric Field distributions $\mid \Re\{E_{x}(y,z)\}\mid$ for the MG design for incident angle of $\theta_\mathrm{in}=70^{\circ}$ at $f=25.18\mathrm{GHz}$. The analytical calculation (a) is compared with the full-wave simulation (b). The dashed black vertical lines indicate the metagrating plane ($z=-h$), while the black circles mark a small region around the meta-atom where deviation between the analytical prediction and the full-wave simulations is expected due to the finite size of the printed capacitors \cite{epstein2017unveiling,rabinovich2018analytical}.}. 
\label{Fig:Field_plot}
\end{figure}

To apply the outlined procedure to our MG prototype, we merely need to substitute the relevant parameters into Eqs. \eqref{eq:final_condition} and \eqref{eq:final_load_impedance} and solve for the fabrication-ready design specifications. We chose to use a Rogers RO3003 laminate of thickness $h=60\mathrm{mil}=1.524\mathrm{mm}$ as the substrate for the prototype, featuring 
anisotropic permittivity of $\varepsilon_{2,xx}=\varepsilon_{\mathrm{2},yy}=3.39\varepsilon_0$, $\varepsilon_{2,zz}=3\varepsilon_0$, and loss tangent of $\tan\delta=0.001$. 
The laminate arrived covered with $0.5\mathrm{oz.}$ electrodeposited copper, corresponding to trace thickness of $t=18\mathrm{\mu m}$; to operate well within standard PCB fabrication constraints, we set the copper trace width and separation to $w=s=6\mathrm{mil}=0.1524\mathrm{mm}$ [Fig. \ref{Unit_Cell}(c)]. 

We used these values in Eq. \eqref{eq:final_condition} and solved the equation for $\left|\rho\right|$ graphically, as illustrated in Fig. \ref{Fig:Theta_out_h}; the solutions correspond to valleys in the plot (purple dahsed lines), where $|\rho|$ approaches $0$. The graph revealed that perfect anomalous reflection with $\theta_\mathrm{in}=70^\circ$, $\theta_\mathrm{out}=-10^\circ$, and the chosen $h = 60\mathrm{mil}$, can be obtained at the working frequency $f=25.18\mathrm{GHz}$, implying periodicities along the $y$ and $x$ directions of $\Lambda=10.69\mathrm{mm} {\approx0.9\lambda}$ and $l=1.19\mathrm{mm}=0.1\lambda$, respectively; in terms of the wavelength, the resulting device thickness corresponds to $h\approx\lambda/8$. Invoking Eq. \eqref{eq:final_load_impedance}, the required capacitance can be assessed, proportional to the printed capacitor width 
as discussed therein. Due to the anisotropic dielectric constant of the substrate, we did not use the analytical approximation as is to retrieve $W$ from $C$, but rather conducted a short sweep of the capacitor width around the estimated value, yielding an optimum at $W=1.39\mathrm{mm}$. 

The resulting MG was defined in a commercial full-wave simulation tool (CST Microwave Studio) to verify the analytical model prior to fabrication. In these simulations, realistic losses were accounted for, using the standard copper conductivity value of $\sigma_\mathrm{copper}=5.96\cdot 10^{7}[S/m]$ in addition to the dielectric losses specified above. Figure \ref{Fig:Field_plot} compares the electric field distributions as calculated using the analytical formulation [Fig. \ref{Fig:Field_plot}(a)] and as recorded in simulations [Fig. \ref{Fig:Field_plot}(b)]. As can be observed therein, excellent agreement between theory and simulations is obtained, except for the expected deviation at the close vicinity of the meta-atoms \cite{epstein2017unveiling,rabinovich2018analytical}.
Overall, these MG simulations imply that the final PCB layout should exhibit extremely high efficiency, with $98.3\%$ of the incident power being anomalously reflected towards $\theta_\mathrm{out}$, limited only by minor absorption by the copper and laminate of $1.7\%$.


\begin{figure}[t]
\centering
\includegraphics[width=8cm]{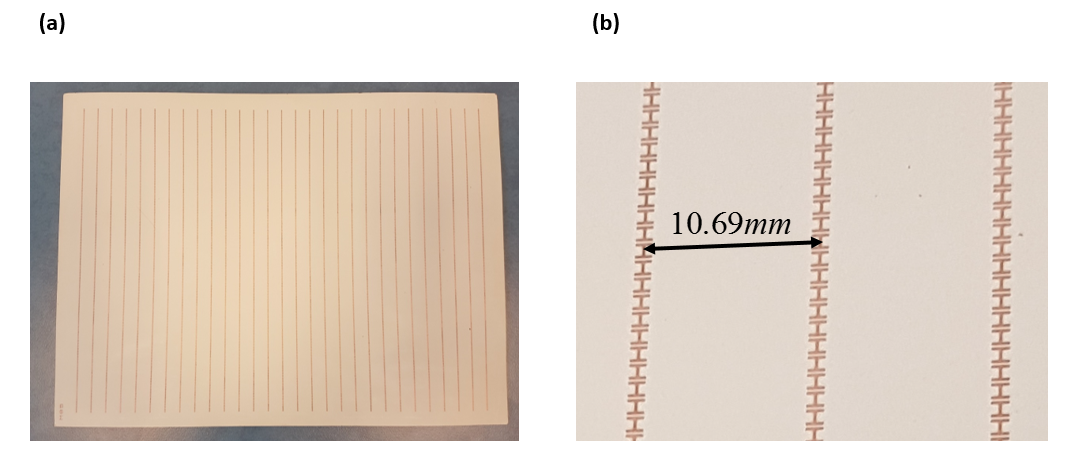}
\caption{(a) The fabricated MG prototype. (b) Zoom in on three lateral periods of the MG.}
\label{Fig:MG_board}
\end{figure}

\begin{figure}[t]
\centering
\includegraphics[width=3.0in]{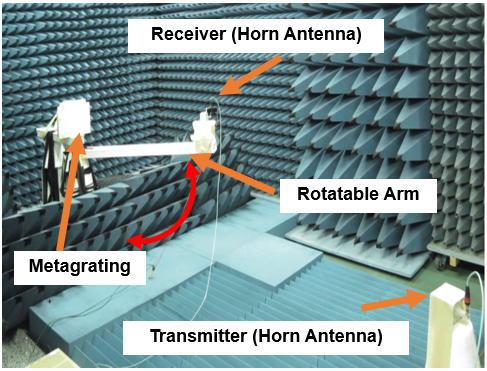}
\caption{Experimental setup assembled in an anechoic chamber, consisting of a transmitter and reciever standard horn antennas, the MG prototype board and a rotatable arm. The transmitter antenna and the MG are fixed in position while the reciever antenna rotates and records the scattered power at each angle.}
\label{Fig:Setup}
\end{figure}

In order to verify the design and characterize the prototype experimentally, a $9''\times12''$ MG was fabricated by \textit{PCB Technologies Ltd.}, Migdal Ha'Emek, Israel; the resulting board is shown in Fig. \ref{Fig:MG_board}. 
To evaluate the device performance, we have assembled an experimental setup in an anechoic chamber in Rafael, Advanced Defense Systems, Ltd. (Fig. \ref{Fig:Setup}), with the MG placed on a stand, and a pair of standard horn antennas (Schwarzbeck, BBHA9170) for excitation and measurement.
The transmitter antenna was aligned at $\theta_\mathrm{in}=70^{\circ}$ relative to the MG, illuminating it from a distance (approximately $2\mathrm{m}\approx168\lambda$); the receiving horn was mounted on a plastic arm connected to a rotatable stage, recording scattered signals approximately $1.2\mathrm{m}\approx100\lambda$ away from the MG. Therefore, by rotating the receiver in angular steps of $\Delta\theta=0.2^{\circ}$ while keeping the transmitter and the MG board fixed and aligned, we could measure the power scattered towards the region $-90^{\circ}<\theta<90^{\circ}$. For each angle, the reflected power was measured as a function of frequency, covering the spectral range of the horn antennas $15\mathrm{GHz}-40\mathrm{GHz}$ in steps of $\Delta f=30\mathrm{MHz}$. Overall, the experiment retrieves the frequency-resolved (bistatic) scattering pattern of the MG when illuminated from $\theta_\mathrm{in}$, denoted as $P_{\mathrm{MG},\theta_\mathrm{in}}\left(f,\theta\right)$. For reference, we have repeated the experiment with the MG replaced by a very thin metallic plate (made of aluminium) with the same dimensions in the $xy$ plane; correspondingly, the recorded scattering patterns are denoted by $P_{\mathrm{Metal},\theta_\mathrm{in}}\left(f,\theta\right)$.

\section{Results and discussion}
\label{sec:results}
\subsection{Anomalous reflection}
\label{subsec:Coupling}
Our first goal is to verify that the fabricated prototype indeed implements the desirable functionality. To this end, it is required to process the raw data as to retrieve the portion of power coupled to the various FB modes scattered off the MG. As denoted in Section \ref{sec:introduction}, in order to evaluate the coupling efficiency to the individual modes as a function of the frequency, we utilize an original beam-integration approach. 
According to this scheme, for each of the considered frequencies, we first identify the beams corresponding to the various propagating modes. This is achieved by looking for peaks of scattered power around the modal propagation angles, as predicted by the FB theorem, namely \cite{rabinovich2018analytical} 
\begin{equation}
\label{eq:Mode_selection}
\begin{aligned}
\sin \theta_{\mathrm{out}}^{\left(m\right)}=\sin\theta_{\mathrm{in}}+\frac{2\pi m}{k_{1}\Lambda},   m=0,\pm1,\pm2...
\end{aligned}
\end{equation}
where the superscript $\left(m\right)$ denotes the $m$th FB mode. 

\begin{figure}[t]
\centering
\includegraphics[width=3.0in]{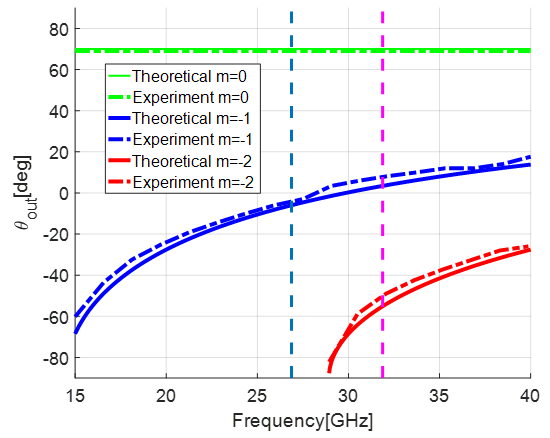}
\caption{The angles $\theta$ where the peak scattered power is observed for each of the propagating modes, as a function of frequency, for excitation from $\theta_\mathrm{in}=70^\circ$. The theoretical predictions (solid lines) following Eq. \eqref{eq:Mode_selection} are compared with the experimentally recorded angles (dash-dotted lines) for the $m=0$ mode (green), the $m=-1$ mode (blue), and the $m=-2$ mode (red). Cyan and magenta vertical dashed lines denote, respectively, the frequencies where optimal anomalous reflection (Section \ref{subsec:Coupling}) and beam splitting (Section \ref{subsec:functionalities}) are observed experimentally.}
\label{Prism_70}
\end{figure}

Figure \ref{Prism_70} presents these angles for the MG illuminated from $\theta_\mathrm{in}=70^\circ$ across the frequency range considered in the experiment, comparing the theoretical predictions of Eq. \eqref{eq:Mode_selection}
to the modal peaks identified in the measured data; it is apparent that the two agree very well.
As can be observed in the figure, although the modal indices can take arbitrary integer values, for the considered angle of incidence only the $m=0$ and $m=-1$ FB modes are propagating for all the frequencies in the relevant range (in consistency with Section \ref{sec:theory}), whereas the $m=-2$ mode becomes propagating only at high frequencies
; for higher order modes, $\theta_\mathrm{out}^{\left(m\right)}$ of Eq. \eqref{eq:Mode_selection} will not be purely real, indicating their evanescent nature.


\begin{figure*}[t]
\centering
\centerline{\includegraphics[width=7in]{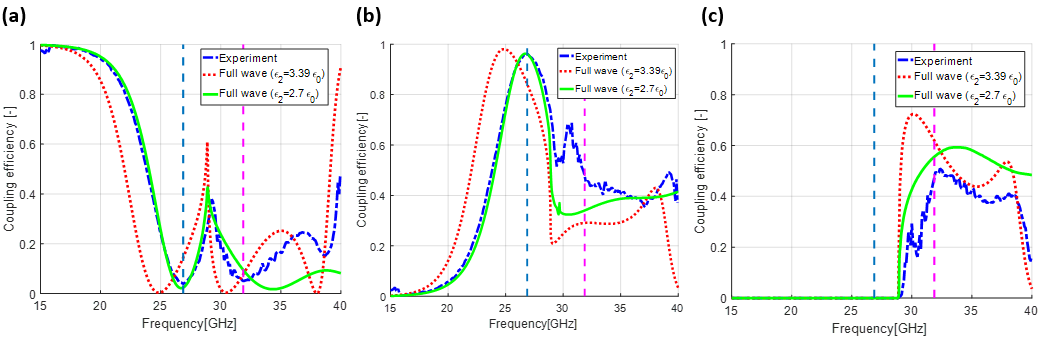}}
\caption{Coupling efficiency as a function of frequency to the (a) $m=0$ (b) $m=-1$, and (c) $m=-2$ FB modes, for $\theta_\mathrm{in}=70^\circ$. The experimental results (dash-dotted blue line) are compared to the results obtained via full-wave simulation with the designed parameters, i.e. $\varepsilon_{r,xx}=3.39, \varepsilon_{r,yy}=3.39, \varepsilon_{r,zz}=3$ (dotted red line), and those from full-wave simulation with the modified effective dielectric constant $\varepsilon_{r}=2.7$ (green solid line). Cyan and magenta vertical dashed lines denote, respectively, the frequencies where optimal anomalous reflection (Section \ref{subsec:Coupling}) and beam splitting (Section \ref{subsec:functionalities}) are observed experimentally.}
\label{Theta_in_70_m}
\end{figure*}

\begin{figure}[t]
\centering
\includegraphics[width=3.0in]{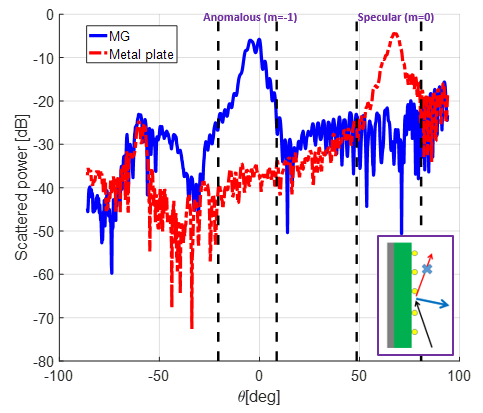}
\caption{Experimentally recorded scattering patterns for the designated angle of incidence $\theta_\mathrm{in}=70^\circ$, at the optimal operation frequency $f=26.87\mathrm{GHz}$ (cyan vertical lines in Figs. \ref{Prism_70} and \ref{Theta_in_70_m}). The received power as a function of $\theta$ as obtained for the MG board (solid blue) is compared to the profile obtained for the reference metallic plate (dash-dotted red) at the optimal frequency, clearly showing the expected suppression of the specular reflection and the efficient coupling to the anomalous reflection $m=-1$ FB mode. \textbf{Inset:} schematic description of the main scattering phenomena.}
\label{Scattering_Pattern}
\end{figure}

Next, we define the beam boundaries for each mode at these angles around the beam peak where the scattered power drops below $-20 \mathrm{dB}$ of the maximum; we denote these angles by $\theta_-^{\left(m\right)}$ and $\theta_+^{\left(m\right)}$ for the $m$th-order mode. 
Subsequently, we integrate the power over the angular range associated with the beam, and divide it by the total power scattered to all the FB modes to estimate the fraction of power coupled to each of the relevant modes. In other words, considering an excitation from $\theta_\mathrm{in}$, the coupling efficiency to the $m$th FB mode can be assessed, as a function of frequency, via
\begin{equation}
\label{eq:coupling efficiency}
\begin{aligned}
\eta_{\mathrm{cpl},\theta_\mathrm{in}}^{(m)}(f)=\frac{\displaystyle\int_{\theta_-^{(m)}}^{\theta_+^{(m)}}P_{\mathrm{MG},\theta_\mathrm{in}}(f,\theta)d\theta}{\displaystyle\sum_{n}\displaystyle\int_{\theta_-^{(n)}}^{\theta_+^{(n)}}P_{\mathrm{MG},\theta_\mathrm{in}}(f,\theta)d\theta},
\end{aligned}
\end{equation}
where, as mentioned above, $P_{\mathrm{MG},\theta_\mathrm{in}}(f,\theta)$ is the recorded scattering pattern for the MG when illuminated from $\theta_\mathrm{in}$ at the frequency $f$, and the summation in the denominator is over all the \emph{propagating} modes in the tested scenario, according to the FB theorem.
It should be noted that although various definitions for the beam boundaries can be considered, e.g., where the scattered power drops to $25\mathrm{dB}$ or $30\mathrm{dB}$ below the identified beam peak, we found that these do not affect much the evaluated coupling efficiencies, as long as the definitions are applied consistently to all modes. 

Considering the typical evaluation procedure of the coupling efficiencies in recent experimental work involving anomalous reflection MSs at microwave frequencies \cite{diaz2017generalized,wong2018perfect}, it may be worthwhile to highlight two appealing features associated with the beam-integration approach followed herein. In previous work, the coupling efficiency was commonly assessed by comparing the peak power recorded at the angle of anomalous reflection for an MG (single measurement point) with the peak specular reflection power measured for a metallic plate of the same dimensions, considering the latter as a good measure for the total scattered power [analogous to the denominator in Eq. \eqref{eq:coupling efficiency}]; due to the different observation angles in which these two peaks (anomalous and specular) are received, proper calibration of the results is required, taking into account the difference in effective aperture size \cite{sievenpiper2005forward, wong2018binary}. In contrast, the beam-integration approach utilized herein relies on a large number of measurement points for the evaluation of the coupled power, corresponding to the angular range associated with the $m$th FB mode beam. Thus, this evaluation method is expected to be more robust to noise effects, averaging the measured data through the integration process [Eq. \eqref{eq:coupling efficiency}]. Using this scheme has an additional positive side effect, as when the overall integrated beam power is considered, effective aperture size differences are inherently taken into account \cite{noteEffectiveAperture}, rendering explicit calibration unnecessary. This can be an advantage, since such a calibration usually requires certain assumptions on the far-field nature of the measurement setup, which are not always easily met.

Applying the beam-integration methodology to the scattering patterns recorded for the MG upon illumination from the designated angle of incidence $\theta_\mathrm{in}=70^\circ$ yields the results presented in Fig. \ref{Theta_in_70_m} (dash-dotted blue curves), where the modal coupling efficiencies for the $m=0$, $m=-1$, and $m=-2$ modes are plotted as a function of the excitation frequency (the rest of the FB modes for this scenario are evanescent).
First, we inspect the coupling efficiency to the anomalous reflection mode $m=-1$ [Fig. \ref{Theta_in_70_m}(b)]. The measured efficiency reaches a near-unity value of $96.11\%$, peaking at $f=26.87\mathrm{GHz}$ (vertical dashed cyan line), which we consider as the optimal operating frequency of the fabricated prototype from now on.
The performance of the MG as an anomalous reflector is further verified by the corresponding scattering patterns recorded at this frequency, presented in Fig. \ref{Scattering_Pattern}. 
The beam boundaries for the $m=-1$ and $m=0$ modes as identified by the automated data analysis code appear as black dashed lines around the predicted angles (see vertical dashed cyan line in Fig. \ref{Prism_70}), clearly marking the dominant scattering from the MG (solid blue) and a metallic plate of the same dimensions (dash-dotted red). As expected, the specular reflection from the MG is highly suppressed with respect to the reference metallic plate, with the scattered power predominantly coupled to the anomalous reflection mode. Indeed, these findings experimentally confirm that the analytically designed PCB MG prototype, featuring a single polarizable particle governed by a single degree of freedom, implements a highly efficient wide-angle anomalous reflection. In addition, as predicted by previous work \cite{ra2017meta,epstein2017unveiling,wong2018perfect,rabinovich2018analytical,popov2018controlling}, the frequency response of Fig. \ref{Theta_in_70_m} reveals a moderate $90\%$ ($-0.5 \mathrm{dB}$) bandwidth of about $8\%$ for the coupling efficiency.

A closer look into Fig. \ref{Theta_in_70_m}(b) reveals that the optimal operating frequency found in the experiment deviates by $6.7\%$ from the design frequency denoted in Section \ref{sec:theory}, as can also be observed by comparison to the coupling efficiency trend of the simulated design (dotted red curve). 
Besides minor fabrication errors and transmitter alignment difficulties, which could somewhat contribute to such a discrepancy, we associate this frequency shift mainly with a deviation between the actual laminate permittivity and the one used for the design. 
Indeed, when we consider a modified substrate permittivity of $\varepsilon_2=2.7\varepsilon_0$ [solid green curve in Fig. \ref{Theta_in_70_m}(b)], the simulations
produce a frequency response which is in close agreement with the measured one across most of the frequency range, both in terms of the optimal operating frequency and the measured bandwidth. 
Moreover, considering the coupling efficiencies to the specular reflection mode [Fig. \ref{Theta_in_70_m}(a)] and to the higher-order ($m=-2$) FB mode [Fig. \ref{Theta_in_70_m}(c)], 
a very good agreement between simulations with the modified permittivity and the measured data is observable as well \cite{noteCouplingEffDeviation70}. 
Hence, considering the expected difference between the manufacturer provided values, measured using a microstrip configuration, and the permittivity values relevant for our scattering scenario, involving plane-wave-like excitations \cite{coonrod2011understanding}, we regard the experimentally evaluated value of $\varepsilon_2=2.7\varepsilon_0$ as the effective laminate permittivity suitable for interpreting the measured scattering patterns.

\subsection{Additional functionalities}
\label{subsec:functionalities}
\begin{figure}[t]
\centering
\includegraphics[width=3.2in]{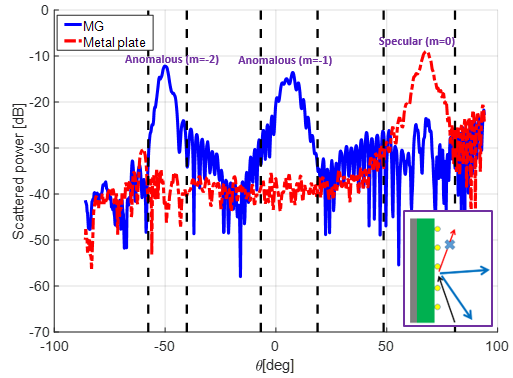}
\caption{Experimentally recorded scattering patterns for the designated angle of incidence $\theta_\mathrm{in}=70^\circ$, at the frequency $f=31.88\mathrm{GHz}$ (magenta vertical lines in Figs. \ref{Prism_70} and \ref{Theta_in_70_m}). The received power as a function of $\theta$ as obtained for the MG board (solid blue) and the reference metallic plate (dash-dotted red) are shown. It is clear that in this working frequency specular reflection is strongly suppressed, while the power splits almost equally between the $m=-1$ and $m=-2$ FB modes, indicating the MG acts as a beam splitter in this frequency. \textbf{Inset:} schematic description of the main scattering phenomena. }
\label{Beam_splitter}
\label{Fig:Beam_splitter}
\end{figure}

As often is the case with experimental work, the measurement setup allows probing the performance of the device under test (DUT) beyond the nominal operating conditions \textit{a priori} considered by design. One such opportunity is provided by the extended frequency range measured throughout the experiment. As can be deduced from Eq. \eqref{eq:Mode_selection} and Fig. \ref{Prism_70}, for frequencies above $28.9 \mathrm{GHz}$, the FB theorem indicates that the second-order mode ($m=-2$) enters the light cone and can be detected in the far field. Scanning this high-frequency region in Fig. \ref{Theta_in_70_m}(a) reveals that the MG features another operating frequency for which specular reflection almost completely vanishes. More specifically, at the frequency $f=31.88 \mathrm{GHz}$ (vertical dashed magenta line in Figs. \ref{Prism_70} and \ref{Theta_in_70_m}) the MG acts as a "perfect" anomalous beam splitter, coupling the wave incident from $\theta_\mathrm{in}=70^\circ$ almost equally to the $m=-1$ ($\theta_\mathrm{out}^{(-1)}=4^\circ$) and $m=-2$ ($\theta_\mathrm{out}^{(-2)}=-55^\circ$) modes [Fig. \ref{Theta_in_70_m}(b) and (c), respectively]. This can be further observed in the corresponding scattering pattern (Fig. \ref{Fig:Beam_splitter}); quantitatively, $\sim47\%$ of the scattered power is coupled to each one of these FB modes, while specular reflection is effectively suppressed, coupling less than $6\%$ of the scattered power. This indicates the potential of MGs to meet the demands of multiband applications, performing different diffraction engineering tasks (different beam steering) at different frequencies.

\begin{figure*}[t]
\centering
\centerline{\includegraphics[width=7in]{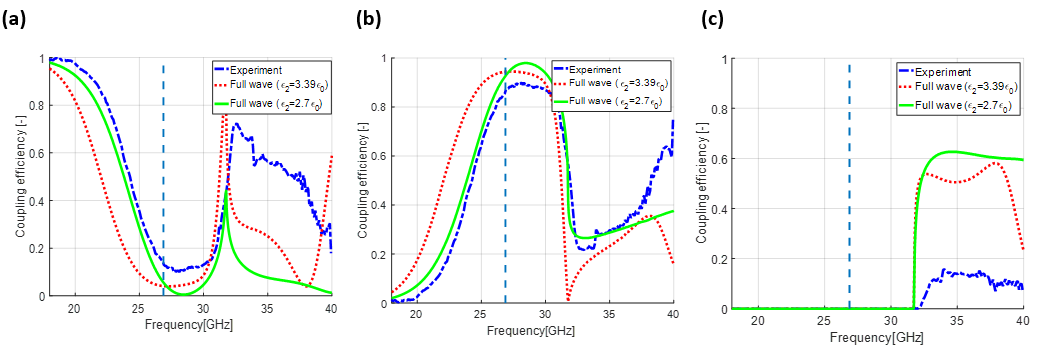}}
\caption{Coupling efficiency as a function of frequency to the (a) $m=0$ (b) $m=-1$, and (c) $m=-2$ FB modes, for $\theta_\mathrm{in}=50^\circ$. The experimental results (dash-dotted blue line) are compared to the results obtained via full-wave simulation with the designed parameters, i.e. $\varepsilon_{r,xx}=3.39, \varepsilon_{r,yy}=3.39, \varepsilon_{r,zz}=3$ (dotted red line), and those from full-wave simulation with the modified effective dielectric constant $\varepsilon_{r}=2.7$ (green solid line). The cyan vertical dashed line denotes the optimal operating frequency observed experimentally for the designated angle of incidence (Section \ref{subsec:Coupling}), $f=26.87 \mathrm{GHz}$.}
\label{Fig:Theta_in_50_m}
\end{figure*}

\begin{figure}[t]
\centering
\includegraphics[width=3.0in]{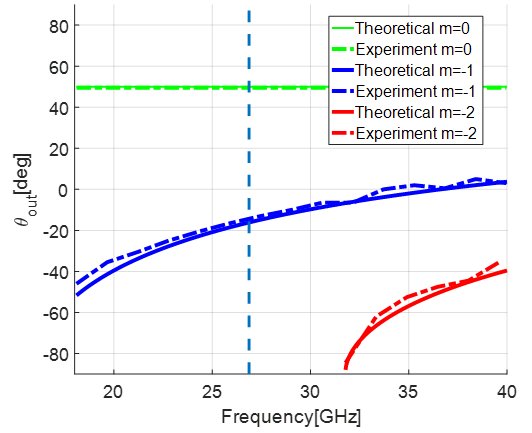}
\caption{The angles $\theta$ where the peak scattered power is observed for each of the propagating modes, as a function of frequency, for excitation from $\theta_\mathrm{in}=50^\circ$. The theoretical predictions (solid lines) following Eq. \eqref{eq:Mode_selection} are compared with the experimentally recorded angles (dash-dotted lines) for the $m=0$ mode (green), the $m=-1$ mode (blue), and the $m=-2$ mode (red). The cyan vertical dashed line denotes the optimal operating frequency observed experimentally for the designated angle of incidence (Section \ref{subsec:Coupling}), $f=26.87 \mathrm{GHz}$.}
\label{Prism_50}
\end{figure}

\begin{figure}[t]
\centering
\includegraphics[width=3.0in]{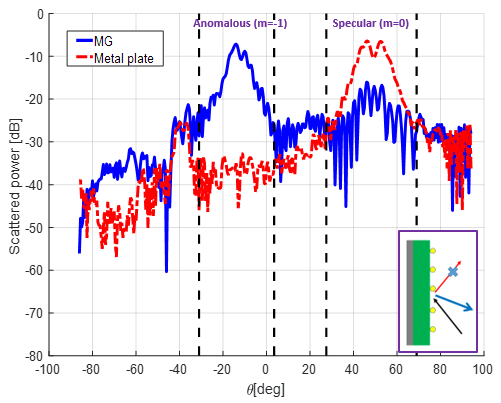}
\caption{Experimentally recorded scattering patterns for $\theta_\mathrm{in}=50^\circ$, at the optimal operation frequency $f=26.87\mathrm{GHz}$ resolved in Section \ref{subsec:Coupling} for the designated angle of incidence (cyan vertical lines in Figs. \ref{Prism_70} and \ref{Theta_in_70_m}). The received power as a function of $\theta$ as obtained for the MG board (solid blue) is compared to the profile obtained for the reference metallic plate (dash-dotted red) at this frequency, indicating that  the MG manages to suppress well the specular reflection even for such a considerable deviation from the designated angle of incidence. \textbf{Inset:} schematic description of the main scattering phenomena.}
\label{Scattering_Pattern_50}
\end{figure}

Another feature single-element MGs inherently exhibit is multichannel reflection. This term, as was utilized recently in the context of MSs \cite{asadchy2017flat} and MGs \cite{ra2018reconfigurable}, refers to simultaneous high-efficiency coupling between waves of different angles of incidence and anomalous reflection. In \cite{asadchy2017flat}, this concept was demonstrated by designing MSs to implement a \emph{discrete} set of reflection channels, harnessing symmetry, reciprocity, and some post-optimization. For the MGs presented herein, on the other hand, \emph{continuous} multichannel reflection is achieved simply due to the non-resonant working point of the meta-atoms, leading to certain insensitivity to deviation from the design parameters \cite{ra2017meta,epstein2017unveiling,wong2018perfect}. In particular, when the angle of incidence deviates from the designated value of $\theta_\mathrm{in}=70^\circ$, scattering is expected to still be predominantly to the anomalous $m=-1$ FB mode, now featuring a different $\theta_\mathrm{out}^{(-1)}$ as per Eq. \eqref{eq:Mode_selection}.

To probe this hypothesis, we have characterized the MG response for an excitation arriving from $\theta_\mathrm{in}=50^\circ$, namely, $20^\circ$ away from the nominal one. The corresponding coupling efficiencies to the anomalous reflection ($m=-1$) FB mode appear in Fig. \ref{Fig:Theta_in_50_m}(b), showing, once more, that good agreement between simulation and measurements is achieved when the modified (effective) permittivity extracted in Section \ref{subsec:Coupling} is used (green solid curve). For the same illumination scenario, Fig. \ref{Prism_50} presents the predicted [Eq. \eqref{eq:Mode_selection}] and measured scattering angles of the various modes, which, again, are in a very good agreement. This figure also highlights one possible reason for the discrepancies between simulation and measurement observed for $f>32 \mathrm{GHz}$ in Fig. \ref{Fig:Theta_in_50_m}(a) and (c), attributed to partial blockage of the transmitting horn under retroreflection $\theta_\mathrm{out}^{(-2)}\approx-\theta_\mathrm{in}=-50^\circ$ (Fig. \ref{Prism_50}), causing underestimation of the coupling to the $m=-2$ mode.

At the operating frequency $f=26.87\mathrm{GHz}$ found in Section \ref{subsec:Coupling} (vertical dashed cyan line), Fig. \ref{Fig:Theta_in_50_m}(b) indeed demonstrates that efficient anomalous reflection is retained even for a substantial deviation from the designated angle of incidence, with coupling efficiencies to the $m=-1$ mode approaching $\sim90\%$ (accompanied by a notable bandwidth of about $15\%$). The corresponding scattering pattern is presented in Fig. \ref{Scattering_Pattern_50}, where it can be clearly seen, by comparison to the scattering from the reference metallic plate, that
the MG greatly suppresses the specular reflection while funnelling the scattered power to the anomalous reflection beam at $\theta_{\mathrm{out}}^{(-1)}=-16^{\circ}$.

\begin{figure}[t]
\centering
\includegraphics[width=3.0in]{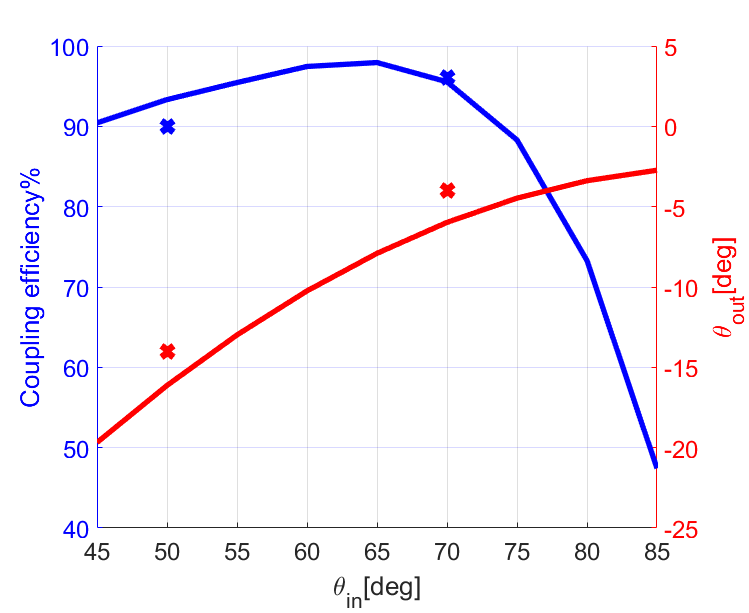}
\caption{The MG as a continuous multichannel reflector for different input angles at the operating frequency $f=26.87\mathrm{GHz}$. The efficiency towards mode $m=-1$ given by full-wave simulation is presented in blue (corresponding to the left y-axis) while the output angle $\theta_\mathrm{out}$ from the analytical calculation [Eq. \eqref{eq:Mode_selection}] is presented in red (corresponding to the right y-axis). The experimentally measured data points corresponding to the incident angles of $50^\circ$ and $70^\circ$ are denoted with X markers.}
\label{Fig:Theta_out_eff_Vs_theta_in}
\end{figure}

Finally, we quantify in Fig. \ref{Fig:Theta_out_eff_Vs_theta_in} the ability of the MG to perform as a continuous multichannel reflector for different angles of incidence at the operating frequency $f=26.87\mathrm{GHz}$. 
Full-wave simulation results (solid lines) indicate that for a broad range of angles $\theta_\mathrm{in}\in\left(45^\circ, 74^\circ\right)$, highly-efficient anomalous reflection (above $90\%$) can be achieved, funnelling the power to the angular range $\theta_\mathrm{out}\in\left(-20^\circ,-5^\circ\right)$. The measured data points for the scenarios probed in the course of this work, $\theta_\mathrm{in}=70^\circ$ and $\theta_\mathrm{in}=50^\circ$ (diamond markers), indicate a very good correspondence with the simulation results, corroborating the concept experimentally.
Reciprocity implies similar multichannel reflection to be observed also for the range $\theta_\mathrm{in}\in\left(-74^\circ, -45^\circ\right)$. Overall, the ability of MGs for multifunctionality, whether in terms of frequency or angle of incidence, is demonstrated.


%
%

\begin{figure}[t]
\begin{subfloat}
\centering
\includegraphics[width=3.0in]{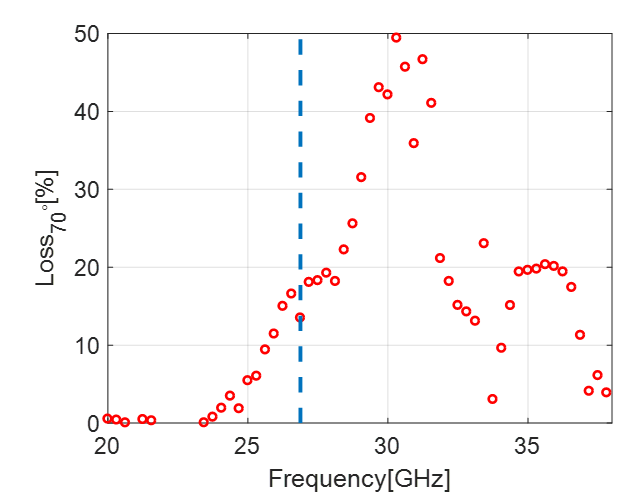}
\end{subfloat}

\begin{subfloat}
\centering
\includegraphics[width=3.0in]{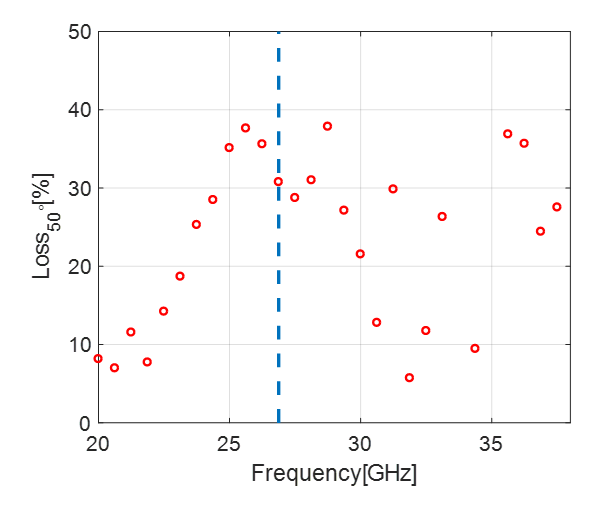}
\end{subfloat}
\caption{The experimentally evaluated loss of the MG according to Eq. (\ref{eq:loss}), for (a) $\mathrm\theta_{in}=70^{\circ}$ and for (b) $\mathrm\theta_{in}=50^{\circ}$. The cyan dashed vertical line denotes the optimal operating frequency defined in Section \ref{subsec:Coupling}.}
\label{Fig:Loss_exp}
\end{figure}



\subsection{Loss mechanisms}
\label{subsec:Loss}

To complete the MG prototype characterization, we seek to quantify its losses. In previous work related to reflecting MSs \cite{wong2018perfect,diaz2017generalized,asadchy2017flat}, this was typically achieved by comparing the anomalous reflection peak power scattered off the MS with the specular reflection peak power scattered off a metallic plate with the same dimensions. Assuming the receiver and transmitter are located in the far field region with respect to the DUT, and measurement noise is sufficiently low, it is expected that for an ideal (namely, lossless and with unitary coupling efficiency) MS, the ratio between these two power maxima would correspond to the ratio between the respective effective aperture sizes of the two measurements (these are different due to the different observation angles \cite{sievenpiper2005forward, wong2018binary}).

Alas, 
even if these assumptions are valid, 
the measurement technique described above can only yield the product of the two contributions to the total anomalous reflection efficiency $\eta_\mathrm{tot}=\eta_\mathrm{scat}\eta_\mathrm{cpl}$, namely, the scattering efficiency ($\eta_\mathrm{scat}$, the fraction of \emph{incident} power that is scattered overall, and not dissipated, for instance), and the the coupling efficiency ($\eta_\mathrm{cpl}$, the fraction of \emph{scattered} power that is coupled to the $m=-1$ mode, and not to spurious modes). In contrast, the beam-integration approach employed in Section \ref{subsec:Coupling} enables decoupling these two contributions, evaluating $\eta_\mathrm{cpl}$ solely based on the MG scattering patterns; this allows verification of the diffraction control properties of the MG, independent of the losses, if exist. 
At the same time, however, the MG measurements alone do not allow quantification of the scattering efficiency.

Consequently, we utilize the scattering patterns measured for the reference metallic plate (dash-dotted red lines in Figs. \ref{Scattering_Pattern}, \ref{Beam_splitter}, and \ref{Scattering_Pattern_50}) to estimate $\eta_\mathrm{scat}$ for the various configurations, and investigate through it the possible loss mechanisms. In particular, we assume the metallic plate possesses negligible losses, such that the overall (integrated) power scattered off it serves as a good reference to the overall power incident upon the MG (recall that the MG and the metal plate have the same dimensions, and are illuminated from the same angle of incidence). Hence, the scattering efficiency can be estimated via
\begin{equation}
\label{eq:loss}
\begin{aligned}
\eta_{\mathrm{scat},\theta_\mathrm{in}}(f)=\frac{\int_{0}^{\pi}P_{\mathrm{MG},\theta_\mathrm{in}}(f,\theta)d\theta}{\int_{0}^{\pi}P_{\mathrm{Metal},\theta_\mathrm{in}}(f,\theta)d\theta}=1-\mathrm{Loss_{\theta_\mathrm{in}}}(f)
\end{aligned}
\end{equation}
where $\mathrm{Loss_{\theta_\mathrm{in}}}(f)$ is defined as the fraction of incident power that is lost in the sense that it is \emph{not} scattered off the MG, as per the recorded patterns at $f$, thus cannot contribute to the desirable anomalous reflection.

\begin{figure}[t]
\begin{subfloat}
\centering
\includegraphics[width=3.0in]{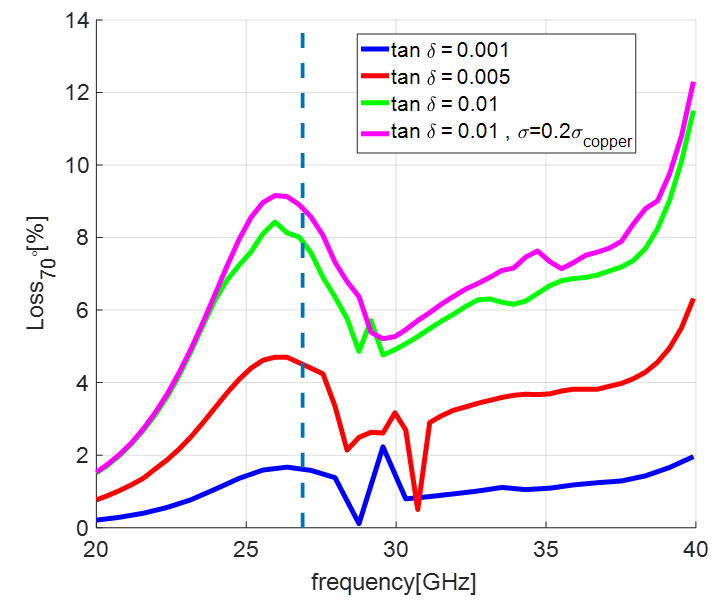}
\end{subfloat}

\begin{subfloat}
\centering
\includegraphics[width=3.0in]{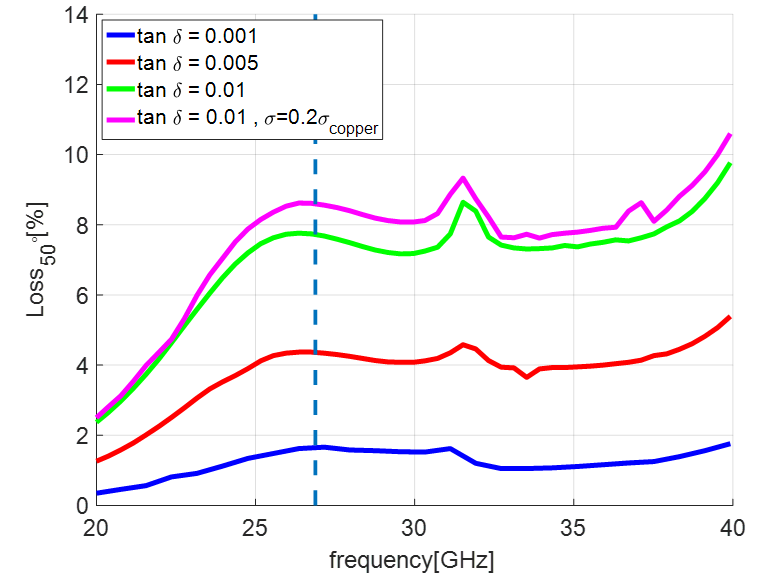}
\end{subfloat}
\caption{MG loss as evaluated from full-wave simulations for various values of dielectric loss and copper conductivity, for angles of incidence (a) $\mathrm\theta_{in}=70^{\circ}$ and (b) $\mathrm\theta_{in}=50^{\circ}$. Increasing substrate loss tangent values are considered, starting from the manufacturer provided $\tan\delta=0.001$ (blue), through $\tan\delta=0.005$ (red), to $\tan\delta=0.01$ (green); in all of these simulations the nominal copper conductivity was used for the wires $\sigma=\sigma_\mathrm{copper}$. To demonstrate that even highly unrealistic loss parameters cannot fully account for the losses deduced experimentally (Fig. \ref{Fig:Loss_exp}), the magenta curve considers a reduced copper conductivity by a factor of $5$, $\sigma=0.2\sigma_\mathrm{copper}$, and loss tangent larger by a factor of $10$, $\tan\delta=0.01$, with respect to the nominal. The cyan dashed vertical line denotes the optimal operating frequency defined in Section \ref{subsec:Coupling}.}
%
\label{Fig:Loss_sim}
\end{figure}

The loss as evaluated by applying Eq. \eqref{eq:loss} to the measured scattering patterns for $\theta_\mathrm{in}=70^{\circ}$ and $\theta_\mathrm{in}=50^{\circ}$ is presented in Figs. \ref{Fig:Loss_exp}(a) and \ref{Fig:Loss_exp}(b), respectively. At the operating frequency $f=26.87 \mathrm{GHz}$ (dashed vertical lines), we observe that $\mathrm{Loss_{70^\circ}}\sim15\%$ and $\mathrm{Loss_{50^\circ}}\sim30\%$. 
In the full-wave simulations presented in Section \ref{sec:theory} for the prototype design, however, we recieved losses in the order of $\sim2\%$. These are contributed by dielectric losses in the substrate of the MG and conduction losses in the copper, realistically accounted for in these simulations.

In order to examine the origin of the increased loss values estimated from the experimental data, we show in Fig. \ref{Fig:Loss_sim} the results of full-wave simulations for enhanced losses, in the dielectric as well as in the conductors. For both angles of incidence considered in the experiment, $\theta_\mathrm{in}=70^\circ$ [Fig. \ref{Fig:Loss_sim}(a)] and $\theta_\mathrm{in}=50^\circ$ [Fig. \ref{Fig:Loss_sim}(b)], we present the predicted $\mathrm{Loss}_{\theta_\mathrm{in}}(f)$ when the loss tangent is gradually increased by an order of magnitude from the nominal $\tan\delta=0.001$ to $\tan\delta=0.01$, and the wire conductivity is reduced by a factor of $5$ with respect to the typical copper conductivity. In all cases considered, it can be observed that even for unrealistically large material loss parameters (e.g., see green curves with $\tan\delta = 0.01$ and $\sigma=0.2\sigma_{copper}$), the dissipated power as predicted by full-wave simulations does not reach the values extracted from the experiments, nor do they exhibit similar trends (Fig. \ref{Fig:Loss_exp}).
Thus, we infer that even if the dissipation in the fabricated MG board is somewhat increased with respect to the manufacturer provided parameters, this alone cannot account fully for the large values of power loss deduced from the experimental data (Fig. \ref{Fig:Loss_exp}); it is evident that an additional mechanism substantially contributes to these observations.

We propose here that the alleged losses presented in Fig. \ref{Fig:Loss_exp} do not correspond to dissipation in the MG, as one may initially suspect, but are mainly a result of edge diffraction. It is well known that finite surfaces, such as the MG and metal plate in our experiment, suffer from edge diffraction effects, stemming from the inevitable distortion of the fields towards the edges and corners (with respect to the  scattering solution of the ideal infinite periodic surface) \cite{ufimtsev2014fundamentals}. Following the common physical optics approximation, for instance, a plane wave propagating in the $\widehat{yz}$ plane impinging upon the DUT, will not scatter only to the $\widehat{yz}$ plane, as in the case of an infinite surface [Fig. \ref{Unit_Cell}(a)], but would also generate scattered waves with non-vanishing $\pm\hat{x}$ wavevector component. As the transmitting and receiving horn antennas in our experimental setup are horizontally aligned with the center of the DUT (Fig. \ref{Fig:Setup}), namely, situated at the plane $x=0$, the edge diffracted power radiating towards $x>0$ and $x<0$ will not be recorded, and might be interpreted as losses. In fact, due to the fact that the MG and metal plate are shorter along the $x$ dimension, these edge diffraction effects are actually expected to be more pronounced along this axis.

Following Eq. \eqref{eq:loss}, however, the estimation of losses in the MG is obtained via comparison to the overall scattering by the reference metal plate. Therefore, if the metallic plate and the MG feature similar edge diffraction characteristics, proportional to the incident power, then the division in Eq. \eqref{eq:loss} should calibrate these effects out. Nonetheless, as we discuss in the following, the edge diffraction from a thin metallic sheet and the MG are fundamentally different. Scattering from a metallic plate is local in essence: the incident power is immediately being reflected (or diffracted) at the point of incidence, and the scattered power is generally proportional to the local field impinging at this point \cite{keller1962geometrical,born2013principles}. MGs, on the other hand, rely on a metal-backed dielectric configuration, which partially guides the incident power in the substrate while reradiating it as reflected waves away from the point of incidence \cite{epstein2016synthesis,diaz2017generalized,asadchy2017flat,kwon2018lossless}. While the latter nonlocal phenomenon does not affect the power balance for an infinite MG (the incident, guided, and reflected modes exchange power periodically to form a uniform response), when finite MGs are considered, the guided power might abruptly meet the edge of the MG before full reradiation has been achieved. If this guided power is substantial, such an event could cause increased edge diffraction with respect to the locally reflecting metallic plate, subsequently interpreted by Eq. \eqref{eq:loss} as loss. 

%

\begin{figure}[t]
\begin{subfloat}
\centering
\includegraphics[width=2.8in]{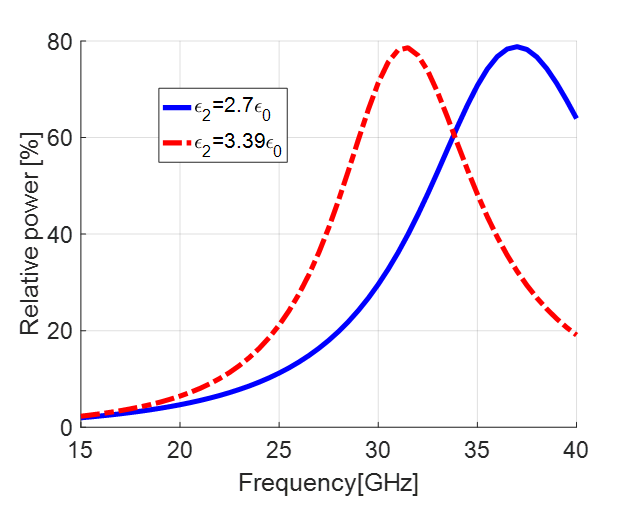}
\end{subfloat}

\begin{subfloat}
\centering
\includegraphics[width=2.9in]{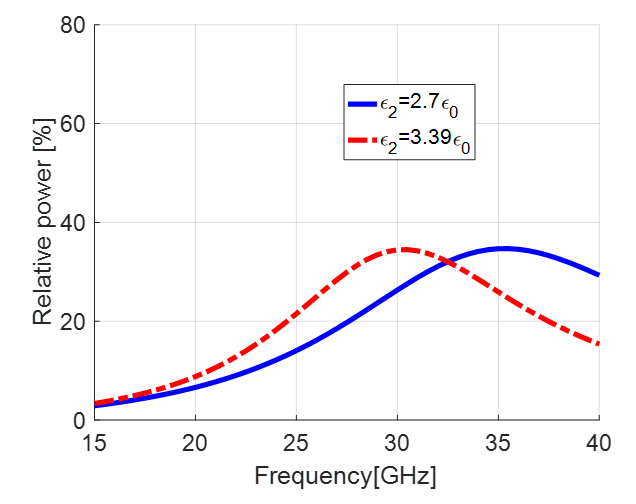}
\end{subfloat}
\caption{The power guided in the dielectric substrate relative to the incident power impinging in a single period, as calculated analytically (see appendix) for (a) $\theta_\mathrm{in}=70^\circ$ and (b) $\theta_\mathrm{in}=50^\circ$. In view of the uncertainty in the substrate permittivities, the two limiting cases of $\varepsilon_2=2.7\varepsilon_0$ (blue solid line) and $\varepsilon_2=3.39\varepsilon_0$ (dash-dotted red line) considered in this paper, are presented.}
\label{Fig:Mode_dielectric}
\end{figure}

\begin{figure}[t]
\centering
\includegraphics[width=2.8in]{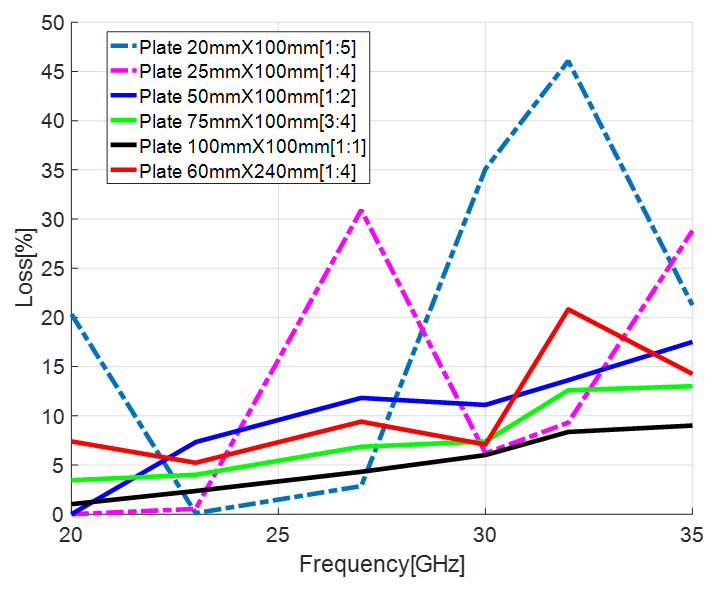}
\caption{The loss associated with finite size metal-backed dielectric substrate when illuminated from $\mathrm \theta_{in}=70^{\circ}$, evaluated from full-wave simulations for different surface areas and aspect ratios of the slab: $20\mathrm{mm}\times100\mathrm{mm}$ (dash-dotted cyan line), $25\mathrm{mm}\times100\mathrm{mm}$ (dash-dotted magenta line), $50\mathrm{mm}\times100\mathrm{mm}$ (solid blue line), $75\mathrm{mm}\times100\mathrm{mm}$ (solid green line), $100\mathrm{mm}\times100\mathrm{mm}$ (solid black line), and $60\mathrm{mm}\times240\mathrm{mm}$ (solid red line). Note that while $L_y$ is fixed on $100\mathrm{mm}$ for most of these configurations, $L_x$ is modified to inspect increasing aspect ratios (denoted in brackets in the plot legend). Finally, the $25\mathrm{mm}\times100\mathrm{mm}$ (dash-dotted magenta line) and the $60\mathrm{mm}\times240\mathrm{mm}$ (solid red line) curves share the same aspect ratio $[1:4]$ but different surface area, indicating that both surface area and aspect ratio are of importance in determining the edge scattering losses as defined herein.}
\label{Fig:Loss_sim_70_finite_size}
\end{figure}

Before providing further support to this hypothesis, two comments are in place. First, we note that the edge diffraction phenomenon as described in the previous paragraph is completely consistent with the full-wave simulation results shown so far, indicating that essentially all of the incident power is coupled to anomalous reflection considering typical dielectric and conductor parameters (Fig. \ref{Fig:Loss_sim}). Indeed, when an infinite periodic MG is considered, no such "scattering losses" are expected (there are no edges to cause edge diffraction). Second, while power guided in the MG substrate is the apparent reason to the "losses" estimated based on the experimental data (Fig. \ref{Fig:Loss_exp}), this phenomenon is far more general, expected to reveal significant edge diffraction differences between any finite-size metal-backed dielectrics and thin metallic plates of the same dimensions. In fact, in order to reduce the complexity of analysis and simulations, we use this simpler configuration to demonstrate said physical effects.

Correspondingly, we calculate analytically the power guided in the substrate (relative to the incident power impinging the surface in a single period) for the scenario of Fig. \ref{Unit_Cell}(a), but excluding the conducting wires (see appendix). More specifically, the reltaive $y$-propagating real power in a dielectric slab of thickness $h=60\mathrm{mil}$ when excited by a plane wave incoming from either $\theta_\mathrm{in}=70^\circ$ or $\theta_\mathrm{in}=50^\circ$ is evaluated as a function of the excitation frequency and plotted in Figs. \ref{Fig:Mode_dielectric}(a) and \ref{Fig:Mode_dielectric}(b), respectively; due to the uncertainty in the substrate permittivity, the results for the two limiting values of $\varepsilon_2=2.7\varepsilon_0$ (solid blue) and $\varepsilon_2=3.39\varepsilon_0$  (dash-dotted red) considered herein are presented. As can be observed, the variation of the losses estimated based on the measurements with frequency and angle of incidence (Fig. \ref{Fig:Loss_exp}) qualitatively correspond to the trend of the relative power guided in the substrate, especially for $\varepsilon_2=3.39\varepsilon_0$. In particular, in both Fig. \ref{Fig:Loss_exp} and Fig. \ref{Fig:Mode_dielectric} the loss peaks corresponding to $\theta_\mathrm{in}=70^\circ$ are higher and narrower than those corresponding to $\theta_\mathrm{in}=50^\circ$, and appear at higher frequencies \cite{noteRelativeGuidedPower}.

Although the agreement is only qualitative (recall that we use a very simplified model here), it clearly ties between the alleged "losses" of Fig. \ref{Fig:Loss_exp} and the guided power causing increased edge scattering for the MG with respect to the metal plate; similar correspondence is observed when considering the power guided in the substrate of the full MG structure, i.e., including the loaded wires (not shown). Combining these observations with the results presented in Fig. \ref{Fig:Loss_sim}, indicating that dielectric and conductor losses cannot on their own explain the frequency response plotted in Fig. \ref{Fig:Loss_exp}, provides supporting evidence to the hypothesis that increased edge diffraction is the main mechanism contributing to the reduced overall scattered power to the $x=0$ plane as recorded for the MG, relative to the reference metal plate. 

In addition to the analysis of the guided power variation with frequency and angle of incidence for the infinite structure, we have also performed a parametric study examining the scattering off a \emph{finite} size surface, this time using full-wave simulations (Fig. \ref{Fig:Loss_sim_70_finite_size}). To enable simulations of large samples approaching the size of the fabricated prototype, we have again considered the simplified metal-backed substrate formation, excluding the conducting wires from the simulated configurations. Denoting the finite dimensions of the structures along the $x$ and $y$ axes as $L_x$ and $L_y$, respectively, we have thus simulated several slabs with different $L_x:L_y$ ratios and various overall surface area $A=L_xL_y$. Each of these slabs was illuminated by a plane wave incoming from $\theta_\mathrm{in}=70^\circ$, and the scattering pattern along the $x=0$ plane was extracted from the simulated 3D scattering pattern to emulate the experimental conditions. Subsequently, the numerical "experiment" was repeated with a corresponding finite-size metallic plate, and the expression in Eq. \eqref{eq:loss} was used to calculate the loss, in consistency with the methodology used to generate Fig. \ref{Fig:Loss_exp}.

This parametric study leads to several important conclusions. First, one can observe that the edge diffraction losses can indeed surmount to significant fractions of the incident power (above $20\%$ for some configurations and frequencies), and they become more severe as the aspect ratio $L_x:L_y$ and the overall area $A$ become smaller. In other words, the off-plane scattering becomes more dominant for the MG when the edges along the $x$ axis are closer to the center of the surface, thus play a more crucial role in the scattering phenomena. However, this also points out a possible way to mitigate this effect, by utilizing square MGs with large areas. Second, it can be observed that for the aspect ratio relevant to the prototype characterized herein, $L_x:L_y=3:4$ (green solid line), the edge scattering losses are predicted to be quite moderate with respect to the ones found experimentally (Fig. \ref{Fig:Loss_exp}). Therefore, we conclude that the transmitting and receiving horns were not perfectly aligned with respect to the MG, and non-uniform illumination of the surface has formed in practice a different effective aspect ratio for the illuminated ares, featuring increased asymmetry leading to increased edge diffraction. Third, although the edge scattering losses as interpreted following Eq. \eqref{eq:loss} can be very significant for certain aspect ratios and surface areas, when the overall integrated 3D power scattered from the dielectric slab and reference metallic plate are compared, simulation results indicate that the overall dissipation in the substrate is minor for the finite structure (not shown), in consistency with the results obtained for the infinite structure (Fig. \ref{Fig:Loss_sim}).

These combined observations imply that the fabricated MG prototype indeed succeeds in implementing a highly-efficient anomalous reflection, not only in the sense of spurious scattering (Section \ref{subsec:Coupling}), but also in the sense of power dissipation. Although it is hard to provide a quantitative limit to the amount of power absorbed in the MG, the investigation presented in this section clearly demonstrates that it is expected to be mild. A more reasonable mechanism to explain the observations of reduced overall scattered power with respect to the reference metallic plate (Fig. \ref{Fig:Loss_exp}) would be edge diffraction effects due to substrate-guided power. Nonetheless, our analysis shows that this effect does not diminish from the MG performance as anomalous reflector: it merely provides some $\hat{x}$ directed momentum to the anomalous reflection mode, without affecting the overall (3D) power balance. Furthermore, simulation results imply that these effects are expected to become less and less dominant as the prototype size increases and the $L_x:L_y$ aspect ratio approaches unity, thus can be mitigated in practical applications.

\section{Conclusion}
To conclude, we have presented a complete demonstration and thorough experimental investigation of a single-element PCB MG for anomalous reflection. The analytical design, fabrication, and testing of a MG prototype was described in detail, verifying its performance as predicted by the analytical model. In addition to exploring the angular and frequency dependence of the prototype, pointing out its potential for implementing multifunctional and multichannel planar reflectors, our report highlights two experimental aspects with relevance to the broad field of reflecting MSs and MGs. First, we have utilized an original beam-integration approach to evaluate the coupling efficiency to the various propagating FB modes. This assessment methodology, which was found to agree very well with full-wave simulations, is expected to be more resilient to measurement noise, and does not require explicit calibration with respect to changes in the effective aperture for different observation angles, non-trivial at times. Second, we shed light on the origin of alleged prototype losses estimated based on the comparison to a reference metal plate. In contrast to the common notions, we showed that it is not plausible that the reduced scattering stems from dissipation in the dielectric substrate and metallic traces; rather, we presented evidence to support a different mechanism to generate similar reductions, namely, edge diffraction due to power guided in the MG substrate. Such a phenomenon, which can be observed in every finite-size devices supporting guided modes, could be relevant to experimental characterization of planar scattering formations in general.

This practical demonstration of efficient analytically-designed MGs forms a crucial step in establishing the viability of this emerging concept for implementing realistic diffraction engineering devices. It indicates that as little as a single degree of freedom in the polarizable element (the printed capacitor width) is sufficient for achieving near-unity coupling of the incident power to a desirable mode, if the proper analytical model is harnessed to this end. These results, together with the more basic observations regarding the experimental characterization of such devices, are expected to contribute to the development of future MG-based components for advanced beam manipulation.

\appendix*
\section{Power guided in the dielectric substrate}

For completeness, we present here the derivation of the analytical expressions for the power guided in a metal-backed dielectric substrate, relative to the impinging power over a single period, used in Section \ref{subsec:Loss} to investigate the expected edge diffraction effects in the corresponding MG structure.
For the configuration under consideration, namely that of Fig. \ref{Unit_Cell}(a) without the conducting wires, the electric field in the dielectric substrate ($-h<z<0$) is given by \cite{rabinovich2018analytical}
\begin{equation}
\label{eq:electric_field_dielectric}
\begin{aligned}
E_{x,2}(y,z)&= \\
-E_\mathrm{in}&\frac{2j\gamma_{0}}{1+j\gamma_{0}\tan{\beta_{0,2}}}\frac{\sin(\beta_{0,2}z)}{\cos(\beta_{0,2}h)}e^{j\beta_{0,1}h}e^{-jk_{t,0}y}
\end{aligned}
\end{equation}
where $E_\mathrm{in}$ is the amplitude of the incident field, $E_{x,1}\left(y,z\right)=E_\mathrm{in}e^{-j\beta_{0,1}z}e^{jk_{t,0}y}$, $k_{t,m}=k_1\sin\theta_\mathrm{in}+\frac{2\pi m}{\Lambda}$, and the parameters $\gamma_m$, $\beta_{m,p}$ are defined in Section \ref{sec:theory} (note that for this scenario, where the periodic wire array is absent, only the zeroth order mode contributes to the scattered fields). Consequently, the $z$ component of the magnetic field is given by 
\begin{equation}
\label{eq:magnetic_field_dielectric}
\begin{aligned}
H_{z,2}(y,z)&= \\
-\frac{k_{t,0}}{k_{2}\eta_{2}}&E_{\mathrm{in}}\frac{2j\gamma_{0}}{1+j\gamma_{0}\tan{\beta_{0,2}}}\frac{\sin(\beta_{0,2}z)}{\cos(\beta_{0,2}h)}e^{j\beta_{0,1}h}e^{-jk_{t,0}y}
\end{aligned}
\end{equation}
Thus, the overall real power travelling in the dielectric in the $y$ direction can be calculated using
\begin{equation}
\label{eq:power_dielectric}
\begin{aligned}
P_{y,2}&(y,z)=\frac{1}{2}\int_{-h}^{0}{\Re\left\{E_{x,2}H_{z,2}^*\right\}}dz=\\
&2\left|E_\mathrm{in}\right|^2\frac{\left|\gamma_{0}\right|^2}{\left| 1+j\gamma_{0}\tan\beta_{0,2}\right|^2}\frac{1}{\cos ^{2}(\beta_{0,2}h)}\\
&\cdot\frac{k_{t,0}}{k_{2}\eta_{2}}\left[\frac{1}{2}h+\frac{1}{4\beta_{0,2}}\sin(-2\beta_{0,2}h)\right],
\end{aligned}
\end{equation}
which is independent of $y$ and $z$.
%
%
%
We normalize this quantity to the power impinging the MG in a single period $\Lambda$, given by
\begin{equation}
\label{power_field_air}
\begin{aligned}
P_{z,1}(y,z)&=\frac{1}{2}\int_{-\Lambda/2}^{\Lambda/2}{\Re\left\{E_{x,1}H_{y,1}^*\right\}}dy=\frac{\beta_{0,1}}{2k_{1}\eta_{1}}\Lambda\left|E_{\mathrm{in}}\right|^{2}
\end{aligned}
\end{equation}
, which is also independent of $y$ and $z$.
Overall, the relative power guided in the dielectric substrate is given by $P_{y,2}/P_{z,1}$, using which the plots presented in Fig. \ref{Fig:Mode_dielectric} were generated.

\begin{acknowledgments}
 This research was supported by the Israel Science Foundation (Grant No. 1540/18). 
The authors also wish to thank Rogers Corporation for providing the laminates used in this study.
\end{acknowledgments}

\vfill

\bibliography{AnalyticalPCBMetagratings_3_9_18}


\end{document}